\documentstyle[epsf]{mn2e}

\title[X-ray spectral evolution in M33 X-8]
{X-ray spectral evolution in the ultraluminous X-ray source M33 X-8}

\author[M.\,J. Middleton, A.\,D. Sutton \& T.\,P. Roberts]
{Matthew J. Middleton$^1$, Andrew D. Sutton$^1$ \& Timothy P. Roberts$^1$\\
$^1$Department of Physics, University of Durham, South Road, Durham
DH1 3LE,
UK\\
}

\pagerange{\pageref{firstpage}--\pageref{lastpage}} \pubyear{2009}

\begin{document}

\topmargin = -0.5cm

\maketitle

\label{firstpage}

\begin{abstract}

The bright ultraluminous X-ray source (ULX), M33 X-8, has been
observed several times by {\it XMM-Newton}, providing us with a rare opportunity to
`flux bin' the spectral data and search for changes in the average
X-ray spectrum with flux level. The aggregated X-ray spectra appear
unlike standard sub-Eddington accretion state spectra which, alongside
the lack of discernible variability at any energy, argues strongly
against conventional two-component, sub-Eddington models.  Although the lack of variability could be consistent with disc-dominated spectra, sub-Eddington disc models are not sufficiently broad to explain the observed spectra.  Fits with a $\sim$ Eddington accretion rate slim disc model are acceptable, but the fits show that the temperature
decreases with flux, contrary to expectations, and this is accompanied
by the appearance of a harder tail to the spectrum.  Applying a
suitable two-component model reveals that the disc becomes cooler and less
advection dominated as the X-ray flux increases, and this is allied to the emergence
of an optically-thick Comptonisation medium.  We present a scenario in which this is explained by the onset of a radiatively-driven wind from the innermost regions of the accretion disc, as M33 X-8 exceeds the Eddington limit.  Furthermore, we argue that the direct evolution of this spectrum with increasing luminosity (and hence radiation pressure) leads naturally to the two-component spectra seen in more luminous ULXs.

\end{abstract}
\begin{keywords}  accretion, accretion discs -- X-rays: binaries -- X-rays: individual: M33 X-8
\end{keywords}

\section{Introduction}

\begin{table*}
\begin{center}
\begin{minipage}{175mm} 
\bigskip
\caption{{\it XMM-Newton\/} observations of M33 X-8.}
\begin{tabular}{l|c|c|c|c|c|c|c}
  \hline

OBSID & obs. date & useful exposure & off-axis angle & ${\it f}_{\rm x}$ & net count rate & total counts & bin\\
 &	& (ks, MOS1) & (arcmin)	& ($\times 10^{-11} \rm ~erg~cm^{-2}~s^{-1}$)	& (ct s$^{-1}$, MOS1)	& \\
   \hline
0102640101 & 2000-08-04 & 8.2 & 1.2 & $1.80 \rightarrow 1.84$ &  1.84 & 30000 & Medium\\
0102640301 & 2000-08-07 & 5.3 & 13.9 & $1.72 \rightarrow 1.79$ &  0.69 &  7500 & Medium\\
0102641001 & 2001-07-08 & 8.7 & 10.5 & $1.63 \rightarrow 1.67$ &  0.88 &  14500 & Low \\
0102642001 & 2001-08-15 & 10.7 & 14.1 & $1.72 \rightarrow 1.78$ & 0.61 &  13000 & Medium\\
0102642101 & 2002-01-25 & 12.2 & 10.7 & $1.75 \rightarrow 1.79$ &  0.96 & 22600 & Medium\\
0102642301 & 2002-01-27 & 12.2 &  8.7 & $1.57 \rightarrow 1.61$ &  1.05 & 24300 &  Low\\
0141980501 & 2003-01-22 & 3.5 & 1.1 & $1.11 \rightarrow 1.16$ &  1.11 &  6800 & Low\\
0141980601 & 2003-01-23 & 12.8 & 14.1 & $1.92 \rightarrow 1.97$ &  0.63 & 16600 & High\\
0141980401 & 2003-01-24 & 7.2 &  12.7 & $1.93 \rightarrow 2.00$ &  0.77 & 12200 & High\\
0141980801 & 2003-02-12 & 12.6 & 1.2 & $1.41 \rightarrow 1.44$ &  1.33 & 26800 & Low\\
0141980101 & 2003-07-11 & 6.1 &  10.6 &$1.49 \rightarrow 1.54$ &  0.88 &  10600 & Low\\
0141980301 & 2003-07-25 & ----- &  & \\
   \hline

\end{tabular}
Notes: The observation date is given in year-month-day format.  The observed flux, $f_{\rm x}$, the background-subtracted net count rate and the total source counts (sum of MOS1 + MOS2) are in the 0.3 -- 10 keV range.  OBSID 0141980301 had no MOS observations and was excluded.
\end{minipage}

\end{center}
\end{table*}

Ultra-luminous X-ray sources (ULXs) are point-like objects with high
($>$10$^{39}$ erg s$^{-1}$) X-ray luminosities that are not associated
with an active galactic nucleus (AGN) or, indeed, the central regions
of a host galaxy (see Miller \& Colbert 2004; Roberts 2007; Gladstone
2011). The nature of these objects has been the subject of much
speculation, with CCD resolution X-ray spectroscopy playing a major
role in advancing our understanding.  Although other missions have
played an important part (e.g. {\it ASCA\/} detection of possible
state transitions in ULXs, Kubota et al. 2001), these results have
predominantly come from the {\it XMM-Newton\/} mission.  Its first
major advance was the detection of a soft excess in the spectra of
many ULXs, with a temperature consistent with that expected for the
inner edge of an accretion disc around an intermediate-mass black hole
(IMBH; e.g. Miller et al. 2003; Miller, Fabian \& Miller 2004).
However, later studies showed that the second, harder component in
these spectra turns over within the {\it XMM-Newton\/} bandpass, and
so appears much cooler and optically thicker than the corresponding
Comptonisation media in Galactic black holes (Stobbart et
al. 2006). This is inconsistent with the identification of a
sub-Eddington state for an IMBH, and more indicative of
super-Eddington accretion onto a stellar-mass black hole (Gladstone, Done \& Roberts 2009).  The apparent
divergence of the spectra of more luminous ULXs into two components
(see Fig. 8 of Gladstone et al. 2009) can be interpreted in terms of
the emergence of a radiatively-driven wind at super-Eddington
accretion rates, with the outflowing material thermalising the
underlying disc emission to produce the soft spectral component as
predicted by e.g. King (2004), Poutanen et al. (2007).  The hard
component is then produced within the photospheric radius, with its
characteristic optically-thick Comptonisation signature either the
result of a thick shroud of Comptonising electrons around the hot
inner disc, or perhaps a change in the opacity of the outer layers of
the hot inner accretion disc itself (Middleton et al. 2011).  Such a
model can explain the startling lack of variability seen in many of
these sources (Heil et al. 2009). In those few cases where large
amounts of variability can be seen, it is likely that the angle of
observation coincides with the edge of the turbulent photosphere,
adding extrinsic variability to the X-ray signal (Middleton et
al. 2011).

Thus it appears that high quality X-ray spectroscopy of ULXs is consistent
with many of the characteristics predicted for super-Eddington
accretion (e.g. Begelman et al. 2006; Poutanen et al. 2007; Mineshige
\& Ohsuga 2011).  However, there is a sub-class of ULXs with
luminosities close to $10^{39} \rm~erg~s^{-1}$, that do not show the
spectral inflection at $\sim 2$ keV indicative of the two-component
spectra in more luminous ULXs, including several objects in the sample
of Gladstone et al. (2009, Fig. 8, top row).  Their spectra are
instead well-described by a single thermal component of emission.
Understanding the nature of this emission and how it fits into the
broader picture of ULX X-ray spectra is important as it may provide a
link between mass accretion rates at $\sim$ Eddington through to
super- and hyper-Eddington rates.  A prime method for
investigating this is to observe the evolution of X-ray spectra with
luminosity for these objects.  Such studies have been undertaken for a
number of more luminous ULXs on the basis of data from {\it Chandra\/}
(e.g. Roberts et al. 2006), {\it Swift\/} (e.g. Vierdayanti et
al. 2010; Kong et al. 2010) and {\it XMM-Newton\/} (e.g. Feng \&
Kaaret 2006; Feng \& Kaaret 2009), yielding important insights including
the demonstration that the temperature-luminosity relation for the
soft excess is indicative of the presence of a wind-driven
photosphere, rather than a standard accretion disc around an IMBH (Kajava \& Poutanen 2009).
However, spectral evolution studies for less luminous ULXs are less
numerous, and can suffer from moderate data quality (e.g. four objects
described by a simple multi-colour disc model in Kajava \& Poutanen 2009).

An obvious means of increasing our understanding of these objects
is to study the brightest example, M33 X-8.  This is the closest
established ULX, long known as the brightest persistent X-ray source
in the Local Group (Long et al. 1981; Gottwald, Pietsch \& Hasinger
1987).  It is also one of only two ULXs (with M82 X-1) whose flux
regularly exceeds $10^{-11} \rm ~erg~cm^{-2}~s^{-1}$ in the {\it
XMM-Newton\/} bandpass, and so has been the subject of many studies
(e.g. Dubus, Charles \& Long 2004; La Parola et al. 2003; Parmar et
al. 2001; Takano et al. 1994).  It has been observed on several
occasions with the EPIC instruments on {\it XMM-Newton\/} (see
e.g. Foschini et al. 2004; 2006).  However, due to the moderate
statistical quality of individual observations it has not been
possible to constrain changes to the X-ray spectrum between
observations (Weng et al. 2009). Here we present a flux binned
analysis that shows that the X-ray spectrum of M33 X-8 does alter
subtly with luminosity, and has a luminosity dependence that is
consistent with the picture of ULX spectral evolution in the
super-Eddington regime.

\begin{figure*}
\begin{center}
\begin{tabular}{l}
 \epsfxsize=13cm \epsfbox{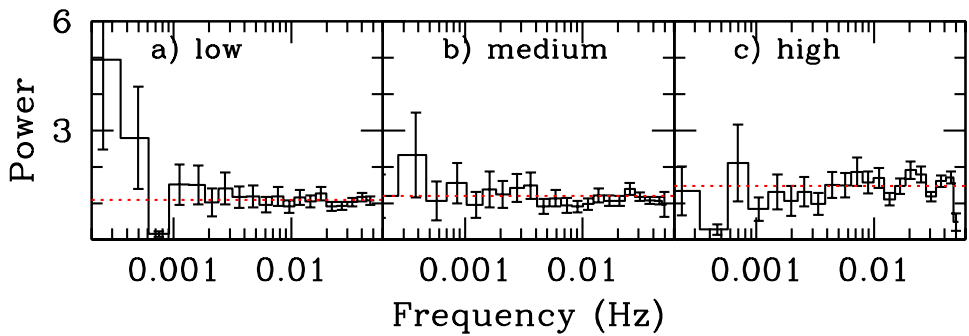}
\end{tabular}
\end{center}
\caption{Average, re-binned (by a geometrical factor of 1.2),
  0.3 -- 10~keV PDS with 1$\sigma$ error bars. The PDS are extracted from
  lightcurves made from the combination of the finessed, background-subtracted, source lightcurves of each flux bin. In practice this has
  meant losing some of the shorter available segments in order to
  obtain a larger frequency bandpass. In the low flux binned data we
  use 401 bins/interval and 4 intervals/frame, in the medium bin, 525
  bins/interval and 4 intervals/frame and in the high flux bin, 431
  bins/interval and 4 intervals/frame. The white noise level is shown
  as the horizontal dashed line and it is clear that there is no constrained variability power density in any bin. This is consistent
  with a model where the spectrum is dominated by stable disc
  emission.}
\label{fig:l}
\end{figure*}

\section{Data analysis}

The EPIC instruments on-board {\it XMM-Newton\/} observed M33 X-8 on
12 occasions of varying durations during the early years of the
mission (see Table 1 for details on each of these)\footnote{New, deep
observations were obtained in the summer of 2010 that are not public
at the time of writing.}. As the source is often not visible in the PN
observations (as it lies off the edge of the chip) we restricted
ourselves to using only the MOS data. The data were initially cleaned
using standard procedures.  Good time interval files were created from
the $> 10$ keV full-field light curves of each observation by removing
periods with a raised background count rate, and used to filter
subsequent data extractions.

Where the source itself appeared in the outer MOS chips we used
circular data extraction apertures of 45 arcsecond radius for both
source and background extractions, and in the three cases where the
source is on-axis (OBSIDs: 0102640101, 0141980501 and 0141980801
respectively) we instead use 30 arcsecond radius regions. We proceeded
to filter the data using {\sc sas v10} and extracted source and
background spectra and lightcurves over standard event patterns ($\leq$
12) and flags (= 0). The background subtracted 0.3 -- 10 keV count
rate is given in Table 1 and suggests that there may be slight pile-up
in OBSIDs 0141980501 and 0141980801 (note that OBSID 0102640101 was
taken in small window mode, and therefore does not suffer from
pile-up at these count rates).  However, inspection of the patterns using the {\sc sas}
tool {\sc epatplot} showed that this is not significant in the MOS1
and only marginally significant for double patterns in the MOS2
observations and as such should not impact a combined spectrum where
the small effect is considerably diluted.

The crude individual spectra appear to be well described by a single
component with a quasi-thermal shape. To obtain flux estimates for
each observation, we model the MOS1 and MOS2 spectral data with a
simple thermal continuum (absorbed {\sc nthcomp}) together with the
{\sc cflux} model in {\sc xspec}, which provides an error range on the
flux based on the data uncertainty. We achieve good or acceptable fits
to each of the individual observation datasets and obtain the absorbed
fluxes given in Table 1. From the observed distribution of fluxes we
break the observations into 3 classes: low, medium and high flux
levels.  Assuming a distance of 817 kpc to M33, consistent with recent
X-ray surveys of the galaxy (e.g. T\"{u}llmann et al. 2011), these
bins correspond to observed luminosities of $< 1.36 \times 10^{39} \rm
~erg~s^{-1}$ (low flux bin), $(1.36 \rightarrow 1.52) \times 10^{39}
\rm ~erg~s^{-1}$ (medium flux bin), and $> 1.52 \times 10^{39} \rm
~erg~s^{-1}$ (high flux bin).  Using the {\sc ftool addspec} we
proceed to co-add the respective datasets (spectra and response
matrices) and obtain a MOS1 and MOS2 dataset for each flux bin.

As our selection of flux bin limits is somewhat arbitrary, we repeat the
model dependent analyses reported in Section 3 on rebinned data, where the
highest flux observation of the low and medium flux bins was promoted into
the next highest bin.  In all cases we find behaviour consistent with the
results reported below, indicating our results are not simply an artefact
of the flux bins used.

The total aggregated counts for each observation are given in Table 1
and, whilst the dispersion in the co-added data will be affected by
the weighted differences between the observations, the data quality
overall is very good for the low and medium flux bins (totals of $\sim
83000$ and 53000 counts respectively). The data quality for the high
flux bin is somewhat poorer ($\sim 29000$ counts), but still a
substantial improvement on most ULX datasets in the {\it XMM-Newton}
archive, and sufficient to constrain the spectra and draw conclusions
on spectral variability in M33 X-8.


\section{flux binned spectroscopy}

An initial inspection of the data using simple empirical models
demonstrated that M33 X-8 does indeed display spectral variability
between the flux binned datasets (which can be trivially confirmed by
inspection of both Tables 2 \& 3, and Figures 2 \& 3).  Here, we focus
on how the spectra evolve with luminosity, and analyse this evolution
in light of a range of physical assumptions and models.

\subsection{Sub-Eddington models}

Several authors have used the high luminosity of ULXs to claim the
presence of an intermediate-mass black hole ($>$500 M$_{\odot}$,
e.g. Kaaret et al. 2003; Miller et al. 2003; Miller, Fabian \& Miller
2004), which must imply sub-Eddington mass accretion rates. In black
hole X-ray binaries such states are characterised by a two-component
model of optically thick disc emission (Shakura \& Sunyaev 1973) and
Comptonisation (see the reviews of McClintock \& Remillard 2006; Done,
Gierli{\'n}ski \& Kubota 2007). We test this assertion by applying a model
containing both of these components together with neutral absorption
and a constant to account for the differences between the detectors
(in {\sc xspec: constant*tbabs*(diskbb+comptt)})\footnote{In these and
all subsequent fits a constant component is utilised.  It is fixed for MOS1,
and the MOS2 value never deviates by more than $\pm 5\%$.
A Galactic column of $1.39 \times 10^{20} \rm ~cm^{-2}$ in the
direction of M33 (Kalberla et al. 2005) is assumed, and used as the
lower bound in fitting the absorption component.}. Whilst Table 2
shows this model provides an acceptable fit to the data in all three
cases, the characteristic temperature of the high energy tail is
$\sim$ 1.2 -- 1.4 keV and so is inconsistent with the high coronal
temperatures seen in sub-Eddington XRB spectra (indicated by the
unbroken power-law continua out to $> 100$ keV, see e.g. McClintock \&
Remillard 2006; although see e.g. Zdziarski et al. 2005 for the
ULX-like behaviour of GRS~1915+105).  We confirm this physical
difference by fixing the plasma temperature at 50~keV in our models,
and measuring the resulting change in fit quality. In all three cases
the fit is poorer with a hot corona, by $\Delta\chi^2$ of 134, 42 and
15 respectively for one more degree of freedom.  This shows that a
cool corona provides a 99.98\% improvement according to the F-test for
the high flux bin, and a substantially more significant improvement
for the other bins, compelling evidence that the corona does not
appear similar to that in standard sub-Eddington states.

\begin{table}
\begin{center}
\begin{minipage}{85mm}
\bigskip
\caption{Best fitting sub-Eddington models.}

\begin{tabular}{l|c|c|c}
\hline

Flux bin & Low & Medium  & High \\
\hline
\multicolumn{4}{|c|}{\sc tbabs*(diskbb+comptt)}\\

$N_{\rm H}$ & 0.069$_{-0.024}^{+0.021}$ &  0.080$\pm 0.006$ &
0.081$\pm 0.011$\\
$kT_{\rm in}$ (keV)  &  0.29$_{-0.03}^{+0.07}$ & 0.66$\pm 0.18$ & 0.67$\pm 0.21$\\
$kT_{\rm comp}$ (keV) & 1.40$_{-0.07}^{+0.08}$ & 1.27$_{-0.10}^{+0.13}$ & 1.40$_{-0.19}^{+0.34}$\\
$\tau$ &  9.18$_{-0.48}^{+0.74}$ & $> 10.14$ & $> 8.33$\\

$\chi^2$ (d.o.f.) & 689.8 (655)  & 654.9 (614) & 521.9  (478)\\

Null P &  0.17 & 0.12  &  0.08 \\
 \\
\multicolumn{4}{|c|}{\sc tbabs*kerrbb} \\

$N_{\rm H}$ & 0.052$\pm 0.003$ & 0.072$\pm 0.003$ & 0.067$_{-0.005}^{+0.003}$ \\
$a$  &  $> 0.999$ & $> 0.971$ & $> 0.991$\\
$\dot{m}$  & 0.50$_{-0.29}^{+0.68}$ & 0.42$_{-0.39}^{+51.98}$ & 0.42$_{-0.36}^{+53.45}$ \\

$\chi^2$ (d.o.f.) & 1080.1  (655)  & 717.3  (614) &  570.6 (478)\\

Null P & 2.0 $\times$10$^{-23}$  & 2.0$\times$10$^{-3}$  &  2.0$\times$10$^{-3}$\\

\hline
\end{tabular}
Notes: Best-fitting parameters for the two sub-Eddington models
used. The units of the foreground column, N$_{\rm H}$, are
10$^{22}$~cm$^{-2}$.  $kT_{\rm in}$ is the temperature of the inner
edge of the accretion disc, and $kT_{\rm comp}$ and $\tau$ are the
temperature and optical depth of the Comptonising medium.  $a$ is the
dimensionless spin parameter and $\dot{m}$ is the `effective' mass
accretion rate of the relativistically-smeared accretion disc (with
the latter in units of 10$^{18}$ g s$^{-1}$), assuming zero torque at
the inner boundary.  The table also shows the values for $\chi^2$ and
the number of degrees of freedom for each model, and the null
hypothesis probability for this model being an acceptable fit to the
data.
 
\end{minipage}
\end{center}
\end{table}

Additionally, the Compton tail in a sub-Eddington state is highly
variable on short timescales providing an unambiguous test for such a
model. We characterise the variability using the excess rms of the 3
-- 10~keV lightcurve in each flux bin (binned to 250 s), i.e. the
variability above the Poisson (white) noise level of the lightcurve
(see Edelson et al. 2002). We find that there is {\bf no} constrained
variability above 3~keV with 3$\sigma$ upper limits of $<$9\%, $<$8\%
and $<$11\% for the three flux bins respectively. This strongly argues
against a two-component sub-Eddington model for the data. Indeed, we
note that the combination of a hot disc (in 2/3 fits), and a cool,
optically thick and invariant corona, demonstrates that this ULX
cannot be described by the cool disc plus power-law continuum toy
model previously used to infer IMBHs in ULXs.

If the data were to be described solely by disc emission then we would
not expect there to be any variability on anything other than the
longest timescales (Wilkinson \& Uttley 2009). As we have substantial
evidence against the two-component sub-Eddington model and are now
interested in the variability behaviour of the emission as a whole, we
extract the Fourier-frequency dependent power density spectra (PDS)
over the full energy bandpass.  Weng et al. (2009) report the lack of
high frequency variability in the individual observations.  Here
however, we can extend this to longer timescales by selecting the
shortest segment of continuous lightcurve and taking integer number of
intervals of this length across the remaining observations. This can
require certain finessing to obtain the maximum amount of available
data and broadest frequency bandpass (as this goes from 1/[longest
available individual segment length] to 1/[2$\times$ bin size]). In
Figure 1 we present the average power density spectrum (PDS -
extracted using the {\sc ftool powspec}) for each flux bin, normalised
to rms$^2$ units following geometrical re-binning with white noise
included. Quite clearly there is no constrained variability over any
of the available frequency bins, consistent with the source having
suppressed red noise as seen in a handful of ULXs (Heil et al. 2009),
and the emission originating in the accretion disc.

However, a thin disc is a poor description of the spectral data as the
Wien tail is not broad enough to fit the high energy emission, leading
to poor fits ($\chi^2$ of 1441.5/658, 931.4/617 and 650.7/481 for an absorbed multi-colour disc blackbody spectrum -- {\sc diskbb} in {\sc xspec} -- fit to each
flux bin respectively). Disc emission that has been smeared by
relativistic effects is considerably broader and so we also fitted the
spectral data with the {\sc xspec} model {\sc constant*tbabs*kerrbb}.  The data and
best fitting models are shown in Figure 2, and the fit parameters are
detailed in Table 2. Even in this extremely relativistically broadened
case we still find the models poorly describe the data, with obvious
excesses above the model at high energies. It would therefore appear
that none of the sub-Eddington models can describe both the broad
shape of the spectra and also the lack of high energy variability.

\begin{figure*}
\begin{center}
\begin{tabular}{l}
 \epsfxsize=13cm \epsfbox{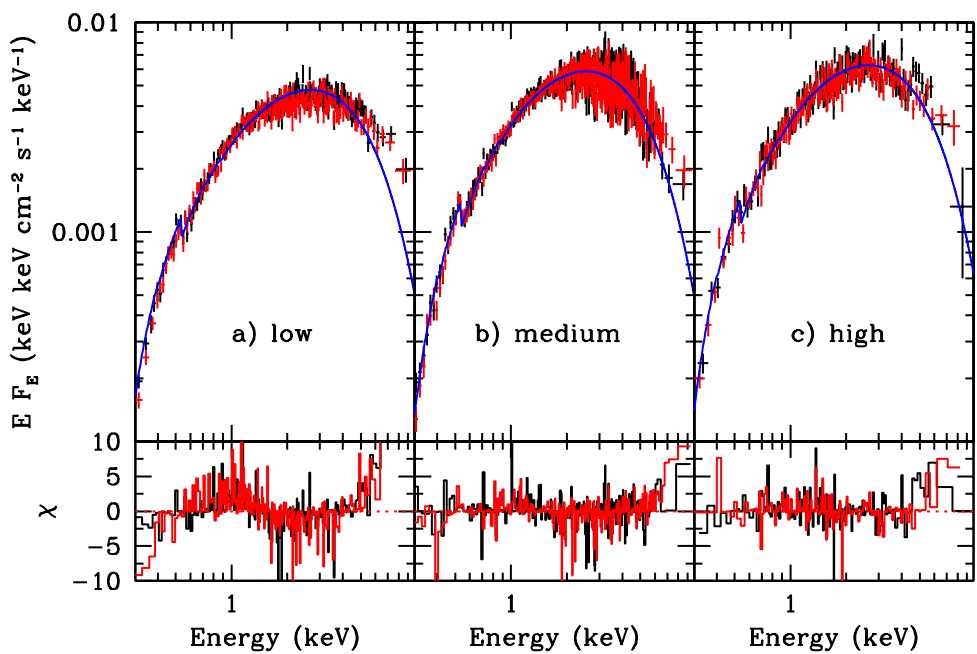}
\end{tabular}
\end{center}
\caption{Flux binned X-ray data for the low, medium and high datasets (MOS1 in
black, MOS2 in red) folded with the best-fitting smeared disc model in
blue ({\sc tbabs*kerrbb}). The residuals to the best fit are shown
below the best-fitting model and demonstrate that, with even a highly
smeared disc, the broad shape of the data is not well matched.}
\label{fig:l}
\end{figure*}

\subsection{Eddington to super-Eddington models}

Examples of Eddington accretion rates (i.e. accretion rates closely
approximating those expected at the Eddington limit for an object)
have been inferred in the outburst of several Galactic X-ray binaries,
for example V404 Cyg ({\.Z}ycki, Done \& Smith 1999), V4641 Sgr
(Revnivtsev et al. 2002) and the neutron star system Cir X-1
(Done \& Gierli{\'n}ski 2003). However, ULXs are observed to be
persistently luminous, which would suggest a closer analogy to a
long-lived outburst system such as GRS~1915+105 (see Remillard \&
McClintock 2006), although this source is also known to undergo
dramatic state changes on short timescales (e.g. Belloni et
al. 2000; Middleton et al. 2006).

Taking a theoretical approach, we would expect that, as the Eddington
limit is neared, the properties of the flow through the disc would
change as advection becomes more important as a method for removing
energy from the flow (Mineshige et al. 2000; Abramowicz et al. 1988).  We would
also expect that, where the flow is highly illuminated, material would be driven off the disc surface by radiation pressure, with the massive outflow forming a photosphere above the disc (see Poutanen et al. 2007).  This
should modify the shape of the continuum emission via absorption,
emission and scattering, producing a fully thermalised, blackbody-like component. Here however, the
spectral shape of M33 X-8 is unlikely to be described solely by a
photosphere covering a hot disc, as the peak temperatures of the
observed spectra are implausibly high for a pure photosphere (c.f. King 2004).

\begin{table}
\begin{center}
\begin{minipage}{85mm} 
\bigskip

\caption{Best fitting Eddington models.}
\begin{tabular}{l|c|c|c}
  \hline

Flux bin & Low & Medium  & High \\
\hline

\multicolumn{4}{|c|}{\sc tbabs*diskpbb}  \\
 
$N_{\rm H}$ &  0.172$_{-0.004}^{+0.011}$ & 0.148$_{-0.010}^{+0.011}$ & 0.151$_{-0.013}^{+0.022}$\\
$kT_{\rm in}$  (keV) & 1.90$_{-0.08}^{+0.13}$ & 1.42 $_{-0.05}^{+0.06}$ & 1.51$_{-0.07}^{+0.16}$\\
$p$  &  0.52$\pm 0.01$ & 0.56$\pm 0.01$ & 0.56$_{-0.02}^{+0.01}$  \\

$\chi^2$ (d.o.f.) & 699.10 (657)  & 667.5 (616) & 532.1  (480)\\

Null P & 0.12  & 0.07  &  0.05  \\

  \\
\multicolumn{4}{|c|} {\sc tbabs*(diskpbb+comptt)} \\

$N_{\rm H}$  & 0.160$\pm 0.010$ & 0.115$_{-0.025}^{+0.023}$ & 0.099$\pm 0.03$ \\
$kT_{\rm in}$ (keV)  & 1.22$\pm 0.04$ & 0.93$\pm 0.02$ & 0.70$_{-0.02}^{+0.03}$ \\
$p$  &  0.54$\pm0.01$ & 0.62$\pm 0.01$ & 0.68$_{-0.01}^{+0.04}$\\
$kT_{\rm comp}$  (keV)& $< 141.76$  & 1.67$_{-0.25}^{+0.56}$ & 1.52$_{-0.19}^{+0.33}$\\
$\tau$ &  $< 5.85$ & 9.63 $_{-2.82}^{+3.23}$ & 10.19$_{-2.01}^{+2.16}$\\

$\chi^2$ (d.o.f.) &  691.0 (654)  &  651.4 (613) & 521.6 (477)\\

Null P &  0.15 & 0.14 & 0.08 \\

   \hline

\end{tabular}
Notes: Best-fitting parameters for the Eddington and/or super-Eddington models. Here $p$ is the dimensionless index of the radial temperature dependence of an advection dominated slim disc (see text), and the other variables are as per Table 2.

\end{minipage}
\end{center}
\end{table}

\begin{figure}
\begin{center}
\begin{tabular}{l}
 \epsfxsize=8cm \epsfbox{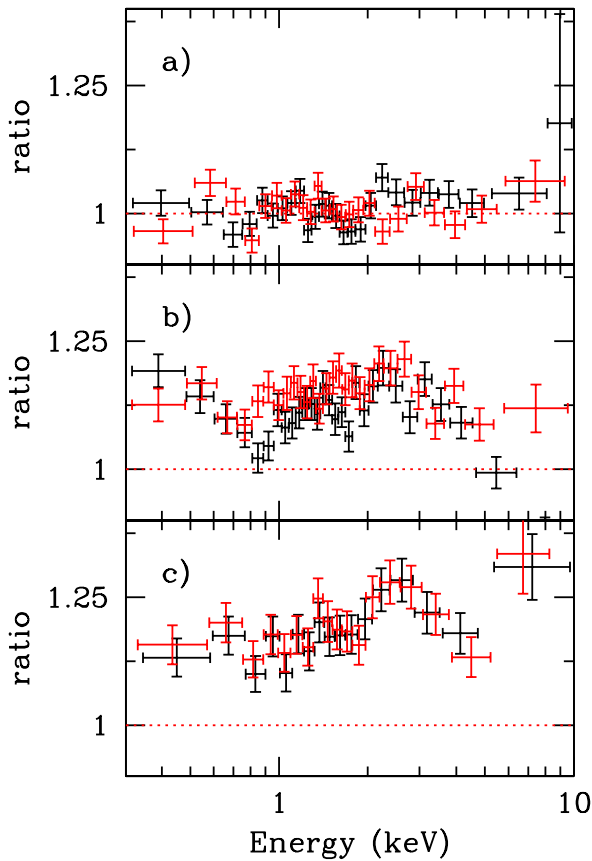}
\end{tabular}
\end{center}
\caption{The ratio of the data in each flux bin to the best fitting
slim disc model ({\sc tbabs*diskpbb}) for the lowest flux bin.  The
panels show the residuals from: a) the low flux bin; b) the medium
flux bin; and c) the high flux bin.  The MOS1 data is shown in black,
and MOS2 in red.  It is clear that, besides the expected increase in
flux, the medium and high flux binned datasets differ in shape with an
apparent deviation in the continuum above $\sim$ 2~keV.}
\label{fig:l}
\end{figure}

We therefore arrive at the model of an advection-dominated Eddington
`slim' disc (Mineshige et al. 2000). Several authors (Foschini et al. 2006;
Weng et al. 2009) have shown this model to be a reasonable description
of the moderate signal-to-noise data of individual observations, with
a reported requirement for an additional soft power-law continuum to high
energies in some cases. Here however, we fit the {\em flux binned} data with
an absorbed $p$-free disc model ({\sc constant*tbabs*diskpbb} in {\sc xspec}), where the index
of the temperature profile $p$ is a free parameter
($T\propto R^{-p}$ for disc temperature $T$ and radius $R$).  Such a model has been used as evidence for slim disc spectra
in ULXs, as a predicted change in the profile from $p = 0.75$ for
standard discs to $p = 0.5$ for slim discs is indeed seen in some
objects (e.g. Vierdayanti et al. 2006; although also see Gladstone et
al. 2009).  The best-fit parameters and their 90\% confidence limits
are shown in Table 3. The model provides a very acceptable description
of the low and medium flux binned datasets, and the peak temperature
is constrained to drop with the increase in flux. The model is a
poorer, although still marginally adequate description of the high
flux data; however its parameters are not significantly different from
the medium flux bin.

Although a slim disc by its very definition does not follow the same
radial temperature profile as the standard thin disc (Shakura \&
Sunyaev 1973), we would not expect the temperature of the inner edge
to drop with increasing flux unless the radius had moved substantially
outwards (contrary to what is seen in Galactic black hole X-ray
binaries).  Similarly, we would not expect the index of the radial temperature dependence $p$ to
increase significantly with flux either. We can see from Table 3 that the temperature
drops by more than 0.3~keV yet the $p$ value increases
significantly. Thus the change here is contrary to what we would
expect for an advection-dominated Eddington flow. However, the decrease in temperature of the peak of the emission could be explained by the inclusion of a flux-dependent cooling mechanism.  In particular, cooling could occur via
the launching of a wind, and Comptonisation in the optically thick
plasma of the outflowing material. In Eddington/super-Eddington flows,
the launching of material is dominated by the radiation pressure
(as the effective gravity goes as 1-L/L$_{Edd}$) and so one
would expect to see its effect increase with luminosity. Hence, in
terms of the spectral evolution, we might expect to see both the disc
cool, and an optically-thick Comptonisation medium emerge, with
increased luminosity.  We illustrate what the data shows in Figure 3,
where we show the spectral residuals for each dataset compared to the
best fitting model for the absorbed $p$-free disc model in the lowest
flux bin (cf. also Table 3).  This shows that, besides the expected
change in the flux level, we also see changes in the shape of the
continuum with flux. In particular, a component with turnover in the 2
-- 3 keV range appears in the medium flux bin spectrum, and appears
more pronounced in the high flux bin.

We attempt to model this by introducing a Compton tail to the model
(using the {\sc comptt} model, such that we employed a {\sc
constant*tbabs*(diskpbb+comptt)} model in {\sc xspec}) with the seed
photons drawn from the soft component in order to provide a physically
limiting case and allow the spectral parameters to be constrained. In
the case of the medium flux binned dataset the fit is notably improved
by this additional component ($\Delta\chi^{2}$ of 16 for 3 d.o.f.s),
and it is also marginally improved in the high flux binned dataset
($\Delta\chi^{2}$$>$ of 11 for 3 d.o.f.s).  Although we would expect
the slim disc emission to dominate at lower luminosities we also fit
this model to the low flux binned data. We obtain only a slightly
improved fit ($\Delta\chi^{2}$$>$ of 8 for 3 d.o.f.s) and, as
expected, the disc dominates the spectrum with only a small
contribution to the higher energy emission made via Comptonisation. In
order to realistically constrain the properties of the best-fitting
two component model, we freeze each best-fitting component in turn and
determine 90\% error limits on the model parameters.  The resulting
fit parameters and errors are presented in Table 3.  They suggest
that, as the luminosity of M33 X-8 increases, the soft component
decreases in temperature and becomes less advection-dominated whilst
the hard component becomes stronger and cooler. A plot of the
evolution of the best-fitting models is shown in Figure 4. As a consistency check we apply these best-fitting models to each of the
individual observations of M33 X-8 in each bin and, in every case, find a
good or acceptable fit to the data.

\begin{figure*}
\begin{center}
\begin{tabular}{l}
 \epsfxsize=13cm \epsfbox{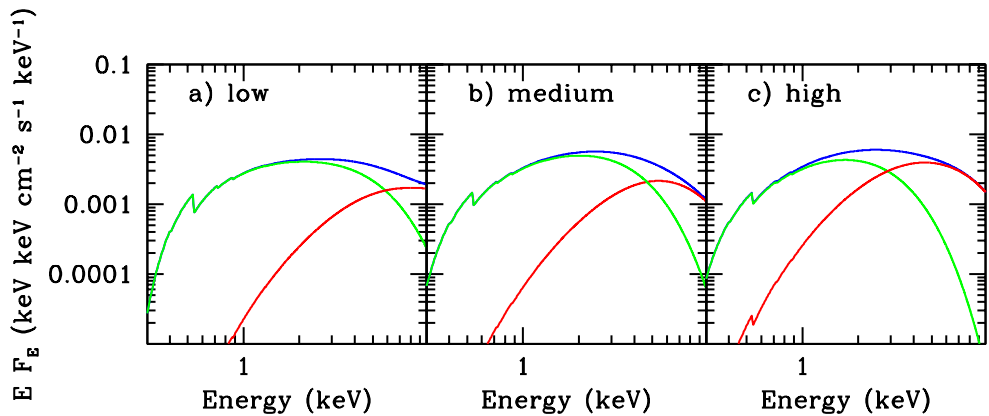}
\end{tabular}
\end{center}
\caption{Best-fitting super-Eddington, two-component models to each
of the flux bins.  The blue line shows the total spectrum, with the
green and red lines showing the disc and Comptonised spectral
components respectively. Although the spectra are broad and can suffer
from model degeneracies, the evolution of a slim disc
spectrum alone from low to medium fluxes appears unphysical, and so this
model presents the most likely physical description of the flux binned
data. The changing properties of the disc and relative contributions
from the two components is consistent with the increased flux leading to
greater wind production, which in turn leads to increased cooling of
the disc and hence an increasingly less advection-dominated solution.
This is naturally accompanied by the emerging optically-thick coronal
component as more material is driven out from the disc.}
\label{fig:l}
\end{figure*}

\section{Discussion \& Conclusion}

Assuming that the population of ULXs contain stellar mass black holes
rather than IMBHs, their spectral evolution should provide the much
sought-after explanation for the changing structure of the accretion flow at and
above the Eddington limit. This requires repeat observations of bright
ULXs at different luminosities. Whilst the relatively large amounts of
data from M33 X-8 have been analysed by other authors (e.g. Weng et
al. 2009, La Parola et al. 2003, etc), it has not previously been flux binned in an
overt attempt to identify spectral evolution. Using this method,
we have been able to test models for the spectral behaviour of this
source.  We argue strongly against a sub-Eddington two-component
description, particularly one describing ULXs as IMBHs, based on the shape of and lack of variability in the
harder spectral component, that results from Comptonisation. The average
PDS of each flux bin further constrained the source's properties and
showed that, across the full energy bandpass, there is no constrained
variability above the white noise at any frequency. Whilst this could
imply that the spectrum is dominated by stable emission from a disc,
we show that neither a standard nor a relativistically broadened thin
(sub-Eddington) disc is a suitable description of the data.

Given the source luminosity of $\sim$1-2$\times$10$^{39}$ erg s$^{-1}$,
Eddington mass accretion rates are easily satisfied by a black hole
mass of $\sim$10M$_{\odot}$, consistent with the masses of Galactic
black holes. We find that an Eddington slim disc provides a good
description of the data with constrained differences in the
temperatures and radial emission profiles of the low and medium
flux binned data. However, these differences are inconsistent with the
expected characteristics of disc accretion, and instead imply the
presence of an additional cooling mechanism.  Subtle changes in the
spectra of the medium and high flux binned states, compared to the low
flux state, are well modelled by including a Comptonised component.

M33 X-8, then, turns out to be a crucial object for furthering our
understanding of accretion at Eddington and super-Eddington rates.  At
its lower fluxes it appears well described by a slim disc model alone;
however as its flux increases, an optically-thick Comptonisation media
becomes more important, in effect `stretching' the disc-like spectrum
by cooling the disc component, and simultaneously providing a harder, upscattered,
spectral component.  This is crucial because it appears to reflect the
initial stages of the divergence of the ULX spectra we see at higher
luminosities, where the spectra are well-modelled by a cool, disc-like
component and a harder, optically-thick Comptonisation component
(Gladstone et al. 2009).  If so, it is likely the result of the
emergence of the expected radiatively-driven wind.

We sketch a toy model of a possible physical scenario to explain both this initial
wind emergence phase, and the higher luminosity photosphere phase in
Figure 5.  In the top panel, the accretion rate is only just entering
the Eddington regime.  Here, the soft component is the outer,
advection-dominated slim disc, which extends down to some radius
(given as R2 in Figure 5). Within this radius the accretion flow is
highly illuminated and we see the emergence of outflowing winds down
to some radius close to the innermost stable circular orbit (ISCO)
beyond which we expect the flow to become highly turbulent (possibly
producing the expected, as yet unobserved, high energy, optically thin
component). The unbound plasma in the wind Comptonises the underlying
hotter disc photons, producing the hard component in the spectrum. As
the luminosity increases, winds may be driven from further out,
whereas closer in, the wind may become increasingly mass loaded as
bound material at lower luminosities may now be lifted from the
`surface' of the disc. This further cools the underlying accretion
flow leading to a cooler peak in the hard spectral component. In
addition, the soft component is now cut off at a larger radius and so
peaks at a lower temperature as well as being less advection dominated (due to being further out in the gravitational potential of the BH). This scenario qualitatively matches the
implied behaviour of M33 X-8.  The key remaining issue is then how
this scenario relates to the spectra seen for more luminous ULXs.  We
emphasise that in this toy model, as the X-ray luminosity increases
we would expect to see the wind being driven from increasingly further
out from the ISCO.  Eventually this would leave little or no disc
emission beyond the outer launching radius, as the very massive wind
would extend over most of the hot (X-ray emitting) regions of the disc and
fully thermalise the underlying disc photons, producing the soft
excess seen in ULX spectra. However, we would also expect to see
`bare' disc emission from within the inner launching radius of the
wind (the photospheric radius) where the local gravity is so low that
the radiation pressure has `blown' the excess material away, leaving an approximate hot thin
disc.  The thermal Comptonisation spectrum then originates either in a
corona of hot electrons tightly bound to the upper layers of the
accretion disc; or perhaps is the spectrum of the accretion disc
itself, as its opacity would alter at the high temperatures within the
photospheric radius.  This scenario is shown in the bottom panel of
Fig. 5, and can explain the spectral properties of the more luminous
ULXs seen to have clear two-component spectra (see Fig. 8 Gladstone et
al. 2009, also Middleton et al. 2011).  In these cases the
wind/photosphere emission is thermally decoupled from the inner disc
emission and so the relative amounts of hard/soft emission are most
likely degenerate in inclination angle and mass accretion rate.

\begin{figure*}
\begin{center}
\begin{tabular}{l}
 \epsfxsize=10cm \epsfbox{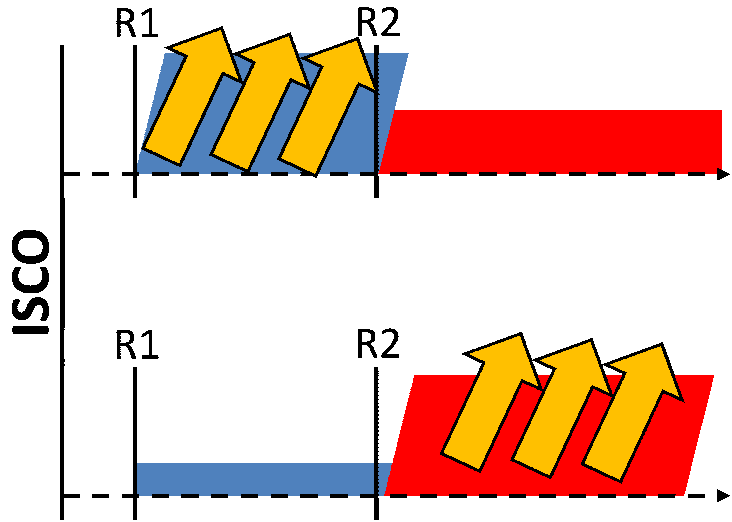}
\end{tabular}
\end{center}
\caption{Our toy model for ULX disc behaviour as the luminosity reaches
(top panel) and far exceeds (bottom panel) its Eddington
luminosity. The top panel shows how, at or near to Eddington, we
expect the majority of the disc to be advection-dominated (slim disc: red), but as the
flow is highly illuminated further in (shown as within the radius R2), we
expect material to be lifted from the surface of the disc by radiation pressure (from the inner
photospheric radius, R1 close to the inner stable circular orbit,
ISCO) which cools the underlying flow. This outflowing, hot material Comptonises the hot disc
photons, producing the hard component in the spectrum, whilst the soft
component originates in the slim disc emission from beyond the outer photospheric
radius (R2). As the mass accretion rate and luminosity increases we would
expect winds to be driven from further out in the disc, and increased mass loading of the wind
closer in. This produces a cooler hot component and a cooler disc
component (as the inner edge of the 'naked' disc emission is now also further
out), as is seen in M33 X-8. The bottom panel then shows what we think may be happening in those
ULXs where the luminosity has increased further, to be substantially super-Eddington. Here,
the radiation pressure is so intense that all the excess material
within R2 has been driven off leaving a bare, hot, thin disc. The
outer launching radius now extends to cover all of the remaining X-ray emitting
disc, producing the soft cool component in the spectrum in a thick photosphere, whilst the hard
component is from the hot bare disc. We suggest that the varying
amounts of each component evident in different ULX spectra (c.f. Gladstone et al. 2009) may be due to degeneracies in inclination
and mass accretion rate between different ULXs.}
\label{fig:l}
\end{figure*}

This model can also consistently explain the associated variability
properties of ULXs as in all cases the spectral components in the
X-ray bandpass are stable (although we predict that at higher energies
there is likely to be an optically thin, highly variable component as
seen in Galactic black hole binaries). However, if our line-of-sight
intercepts the launching region of the photosphere at high
luminosities where the wind is in the outer rather than inner disc,
then we expect this extrinsic variability to lead to the hard
component being highly variable (e.g. NGC 5408 X-1: Middleton et
al. 2011).

Hence, by examining how the X-ray spectral and timing properties of
the nearest ULX vary with source flux, we have shown that it is
consistent with the picture of a source accreting at the threshold of
the super-Eddington regime.  As its luminosity increases, it begins to
betray the signature of a possible outflow from its central regions.
This may be the emergence of the outflow that appears to dominate the
characteristics of ULXs at higher luminosities.  This work shows the
power of considering both spectral and timing data, using the highest
quality datasets available from {\it XMM-Newton\/}.  Clearly, if we
are to develop a deeper understanding of how ULXs work, and
further investigate this picture of objects dominated by a
radiatively-driven wind, obtaining an increased number of similar datasets must be a
priority for this and future missions.

\section{Acknowledgements}

We thank the anonymous referee for their useful suggestions. MM and TR
thank STFC for support in the form of a standard grant, and AS
similarly thanks STFC for support via a PhD studentship. This work is
based on observations obtained with {\it XMM-Newton}, an ESA science
mission with instruments and contributions directly funded by ESA
Member States and NASA.

\label{lastpage}

\end{document}